\documentclass[twocolumn]{aastex62}
\pdfoutput=1

\usepackage{natbib}

\shorttitle{Water-band Photometric Variability of HD 203030B}
\shortauthors{Miles-P\'aez et al.}

\begin{document}

\title{Cloud Atlas: Variability in and out of the Water Band in the Planetary-mass HD 203030B Points to Cloud Sedimentation in Low-gravity L Dwarfs}

\correspondingauthor{Paulo A. Miles-P\'aez}
\email{ppaez@uwo.ca}

\author{Paulo A. Miles-P\'aez}
\affiliation{Department of Physics \& Astronomy and Centre for Planetary Science and Exploration, The University of Western Ontario, London, Ontario N6A 3K7, Canada}
\affiliation{Department of Astronomy/Steward Observatory, The University of Arizona, 933 N. Cherry Avenue, Tucson, AZ 85721, USA}

\author{Stanimir Metchev}
\affiliation{Department of Physics \& Astronomy and Centre for Planetary Science and Exploration, The University of Western Ontario, London, Ontario N6A 3K7, Canada}
\affiliation{Department of Astrophysics, American Museum of Natural History, Central Park West at 79th Street, New York, NY 10024-5192, USA}

\author{D\'aniel Apai}
\affiliation{Department of Astronomy/Steward Observatory, The University of Arizona, 933 N. Cherry Avenue, Tucson, AZ 85721, USA}
\affiliation{Department of Planetary Science/Lunar and Planetary Laboratory, The University of Arizona, 1640 E. University Boulevard, Tucson, AZ 85718, USA}
\affiliation{Earths in Other Solar Systems Team, NASA Nexus for Exoplanet System Science}

\author{Yifan Zhou}
\affiliation{Department of Astronomy/Steward Observatory, The University of Arizona, 933 N. Cherry Avenue, Tucson, AZ 85721, USA}

\author{Elena Manjavacas}
\affil{W. M. Keck Observatory, 65-1120 Mamalahoa Highway, Kamuela, HI 96743, USA}
\affil{Department of Astronomy/Steward Observatory, The University of Arizona, 933 N. Cherry Avenue, Tucson, AZ 85721, USA}

\author{Theodora Karalidi}
\affiliation{ Planetary Sciences Group, Department of Physics, University of Central Florida, 4111 Libra Dr, Orlando, FL, 32816, USAA}

\author{Ben W. P. Lew}
\affiliation{Department of Astronomy/Steward Observatory, The University of Arizona, 933 N. Cherry Avenue, Tucson, AZ 85721, USA}
\affiliation{Department of Planetary Science/Lunar and Planetary Laboratory, The University of Arizona, 1640 E. University Boulevard, Tucson, AZ 85718, USA}

\author{Adam J. Burgasser}
\affiliation{Center for Astrophysics and Space Science, University of California San Diego, La Jolla, CA 92093, USA}

\author{Luigi R. Bedin}
\affiliation{INAF Osservatorio Astronomico di Padova, Vicolo Osservatorio 5, I-35122 Padova, Italy}

\author{Nicolas Cowan}
\affiliation{Department of Earth \& Planetary Sciences and Department of Physics, McGill University, 3550 Rue University, Montr\'eal, Quebec H3A 0E8, Canada}

\author{Patrick J. Lowrance}
\affiliation{IPAC-Spitzer, MC 314-6, California Institute of Technology, Pasadena, CA 91125, USA}

\author{Mark S. Marley}
\affiliation{NASA Ames Research Center, Mail Stop 245-3, Moffett Field, CA 94035, USA}

\author{Jacqueline Radigan}
\affiliation{Utah Valley University, 800 West University Parkway, Orem, UT 84058, USA}

\author{Glenn Schneider}
\affiliation{Department of Astronomy/Steward Observatory, The University of Arizona, 933 N. Cherry Avenue, Tucson, AZ 85721, USA}

\begin{abstract}

We use the Wide Field Camera 3 on the {\sl Hubble Space Telescope} to spectrophotometrically monitor the young L7.5 companion HD~203030B. Our time series reveal photometric variability at 1.27\,$\mu$m and 1.39\,$\mu$m on time scales compatible with rotation. We find a rotation period of $7.5^{+0.6}_{-0.5}$ h: comparable to those observed in other brown dwarfs and planetary-mass companions younger than 300 Myr. We measure variability amplitudes of $1.1\pm0.3\%$ (1.27\,$\mu$m) and $1.7\pm0.4\%$ (1.39\,$\mu$m), and a phase lag of 56$^\circ\pm$28$^\circ$ between the two light curves. We attribute the difference in photometric amplitudes and phases to a patchy cloud layer that is sinking below the level where  water vapor becomes opaque. HD 203030B and the few other known variable young late-L dwarfs are unlike warmer (earlier-type and/or older) L dwarfs, for which variability is much less wavelength-dependent across the 1.1--1.7$\mu$m region. We further suggest that a sinking of the top-most cloud deck below the level where water or carbon monoxide gas become opaque may also explain the often enhanced variability amplitudes of even earlier-type low-gravity L dwarfs.  Because these condensate and gas opacity levels are already well-differentiated in T dwarfs, we do not expect the same variability amplitude enhancement in young vs.\ old T dwarfs.
\end{abstract}


\section{Introduction} \label{sec:intro}

Ultra-cool \citep[$>$M7;][]{reid_etal02} dwarfs contain condensate particles in their atmospheres \citep{1996AA...308L..29T,2001ApJ...556..357A,burrows_etal01} that form a vertically layered structure as a function of condensation temperature \citep{2001ApJ...556..872A,lodders_fegley06}. Inhomogeneities in these layers are responsible for spectrophotometric variability, now commonly observed with long-duration high-precision photometry \citep{artigau_etal09,2014ApJ...793...75R,2014ApJ...782...77B,2015ApJ...799..154M}. 

Clouds are also expected in the atmospheres of giant exoplanets, as these share similar temperatures and chemical compositions as ultra-cool dwarfs \citep{burrows_etal01,2016ApJS..225...10F}. Contrary to most exoplanets, the light of an ultra-cool dwarf can be directly observed by means of photometry and spectroscopy. Thus, ultra-cool dwarfs are excellent laboratories to investigate the physical and chemical processes that take place in their atmospheres, and to develop tools for the atmospheric characterization of exoplanets.

The faint HD 203030B ($J_{MKO}=18.77\pm0.08$ mag) is a young L7.5 companion to the G8V solar analog HD~203030 discovered by \citet{2006ApJ...651.1166M}.  HD~203030B is separated by 11$\farcs$9 \citep[468 AU; {\sl Gaia} DR2 trigonometric parallax is $25.45\pm0.06$ mas;][]{2016A&A...595A...1G} from its primary, and so is easily accessible for spectroscopic characterization. \citet{2017AJ....154..262M} estimated that the system has an age of 30--150~Myr---younger than the 130--400~Myr inferred in the discovery paper---based on low-gravity features in the near-infrared spectrum of the companion, activity indicators of the primary, and the positions of both components on color-magnitude diagrams. The revised age gives HD~203030B a mass of 8--15 $M_{\rm Jup}$ (1$\sigma$) and an effective temperature of $1040\pm50$~K based on evolutionary models  \citep{saumon_marley08, baraffe_etal15}. 

In this work we present near-infrared photometric and spectroscopic monitoring of HD~203030B with the {\sl Hubble Space Telescope} ({\sl HST}; Sec.~\ref{obs}).  We analyze the detected variability in the context of cloud condensation across the L-to-T spectral type  transition.  We draw a parallel to emerging evidence for enhanced photometric variability in young L dwarfs \citep{2015ApJ...799..154M, 2018MNRAS.474.1041V}, and propose a common picture to explain this phenomenon (Sec.~\ref{results}).

\section{Observations and Data Reduction}\label{obs}

We collected photometric and spectroscopic data of HD~203030B as part of the {\sl Cloud Atlas} program (P.I. D. Apai, GO~14241) with the {\sl HST} Wide Field Camera 3 (WFC3) in its near-infrared channel \citep[][]{2010SPIE.7731E..0ZM} and in subarray mode (256$\times$256 pixels). 

\begin{figure*}
\centering
\includegraphics[width=0.98\textwidth]{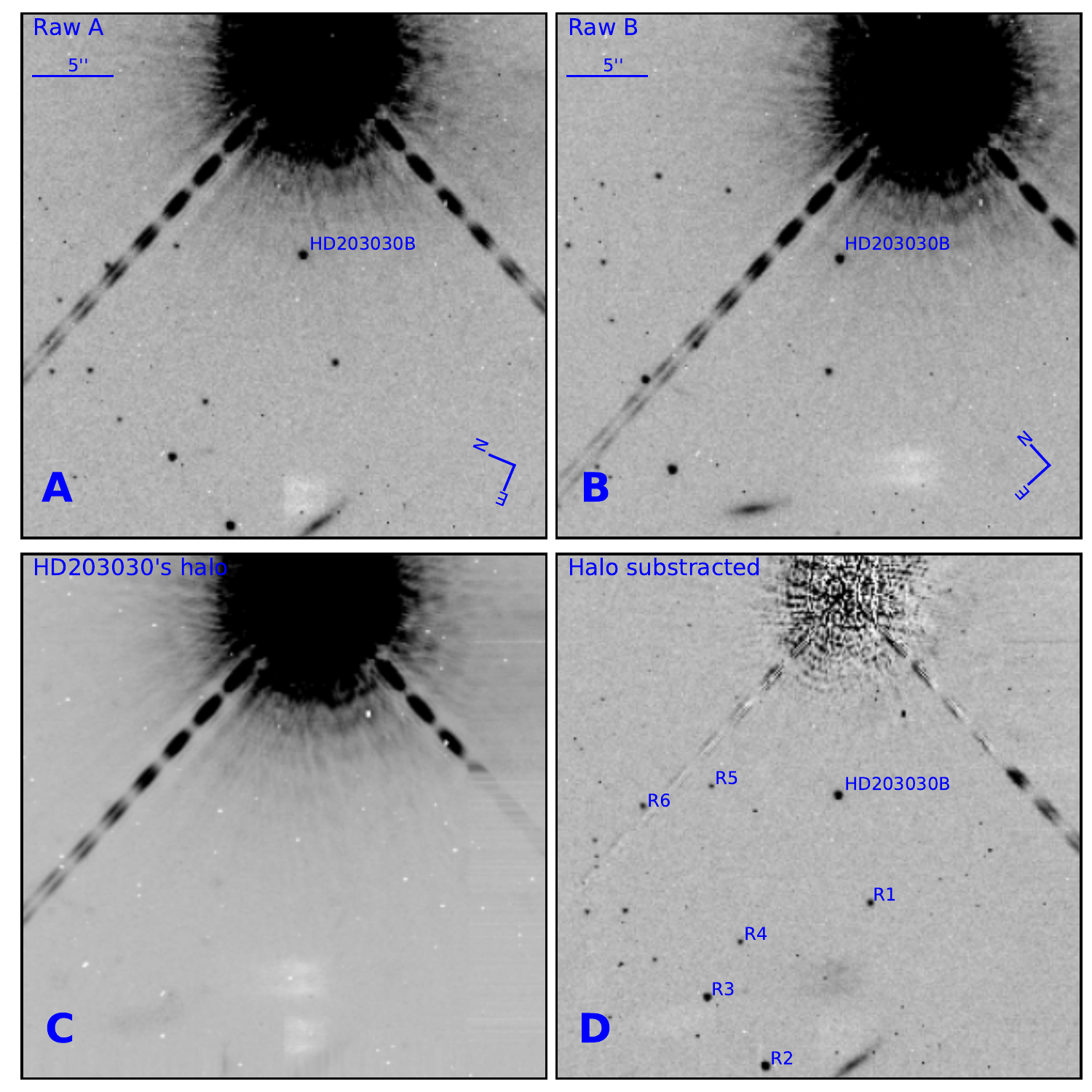}
\caption{WFC3 F127M-band images of HD~203030B at each of the two spacecraft orientations (panels A and B). Panel C shows the primary star and its halo after median-combining all aligned images. The result of subtracting this image from the science images is shown in panel D.  The six reference stars R1--R6 considered for our flux calibration are also shown in the bottom right panel.  Similar data were also collected and analyzed for the F139M filter. \label{im}}
\end{figure*}

\subsection{Near-infrared Photometry \label{sec}}

We obtained near-infrared imaging sequences in the F127M ($\lambda_{\rm central}=1.270\mu$m, ${\rm FWHM} =0.070\mu$m) and F139M ($\lambda_{\rm central}=1.395\mu$m, ${\rm FWHM}=0.070\mu$m) filters over six consecutive orbits on 30 October 2017. We used the SPARS10 sampling mode with individual exposure times of 66.4 s (F127M) and 88.4 s (F139M). At each orbit, data were collected by alternating the F127M and F139M filters every 4 and 5 images, respectively. In total we obtained 78 (F127M) and 84 (F139M) images. The orientation of the spacecraft in orbits 1, 3, and 5 differed by 25$^\circ$ from that in orbits 2, 4, and 6,  following the same procedure as \cite{2016ApJ...818..176Z}. This strategy allowed us to subtract the bright halo of the primary star. Images in these two orientations are shown in panels A and B of Figure \ref{im}.

We measured the flux of HD 203030B from the .flt files generated by the WFC3 pipeline {\sc CALWFC3}. These images are already corrected for non-linearity using an up-the-ramp fit to the flux in non-destructive readouts, and have then been corrected for dark current and flat field effects. We removed bad pixels (flagged with values at 4, 32, 256, or 512) by 1-D interpolation over the nearest neighbors in the same row, as done in previous works by our team \citep{2013ApJ...768..121A,2014ApJ...782...77B,2015ApJ...798L..13Y,2016ApJ...829L..32L,2017Sci...357..683A,2018AJ....155...11M,2018AJ....155..132Z,2019ApJ...875L..15M,2019AJ....157..128Z}. No bad pixels were found inside the photometric aperture of HD~203030B nor the comparison stars. We determined the center of the primary star HD~203030 by fitting lines to its diffraction spikes. We then aligned all images without rotating and median-combined them to obtain a high signal-to-noise image of the HD~203030 point spread function (PSF) and halo. We excluded the upper half of the fluxes at each pixel location in the median combination to mitigate the effect of the diffraction spikes between the observations at the two roll angles.  We finally subtracted that from all individual science images. The median-combined image and an example of a PSF-subtracted image are shown in panels C and D of Figure \ref{im}.

WFC3 is known to suffer from charge trapping within the detector pixels, which leads to a ``ramp'' shape in time-series WFC3 photometry \citep{2012ApJ...747...35B}. This systematic can be corrected by modeling the instrument response \citep{2017AJ....153..243Z}, with an analytic function fit to a non-variable star in the image \citep{2013ApJ...768..121A,2016ApJ...829L..32L}, or via differential photometry. We opted for the latter method as we had several reference stars in our field of view; we detail the approach below.  As a verification, we independently corrected the ramp effect in our photometry with the deterministic instrument response model of \cite{2017AJ....153..243Z}, and found fully consistent results.

\begin{figure}
\centering
\includegraphics[width=0.49\textwidth]{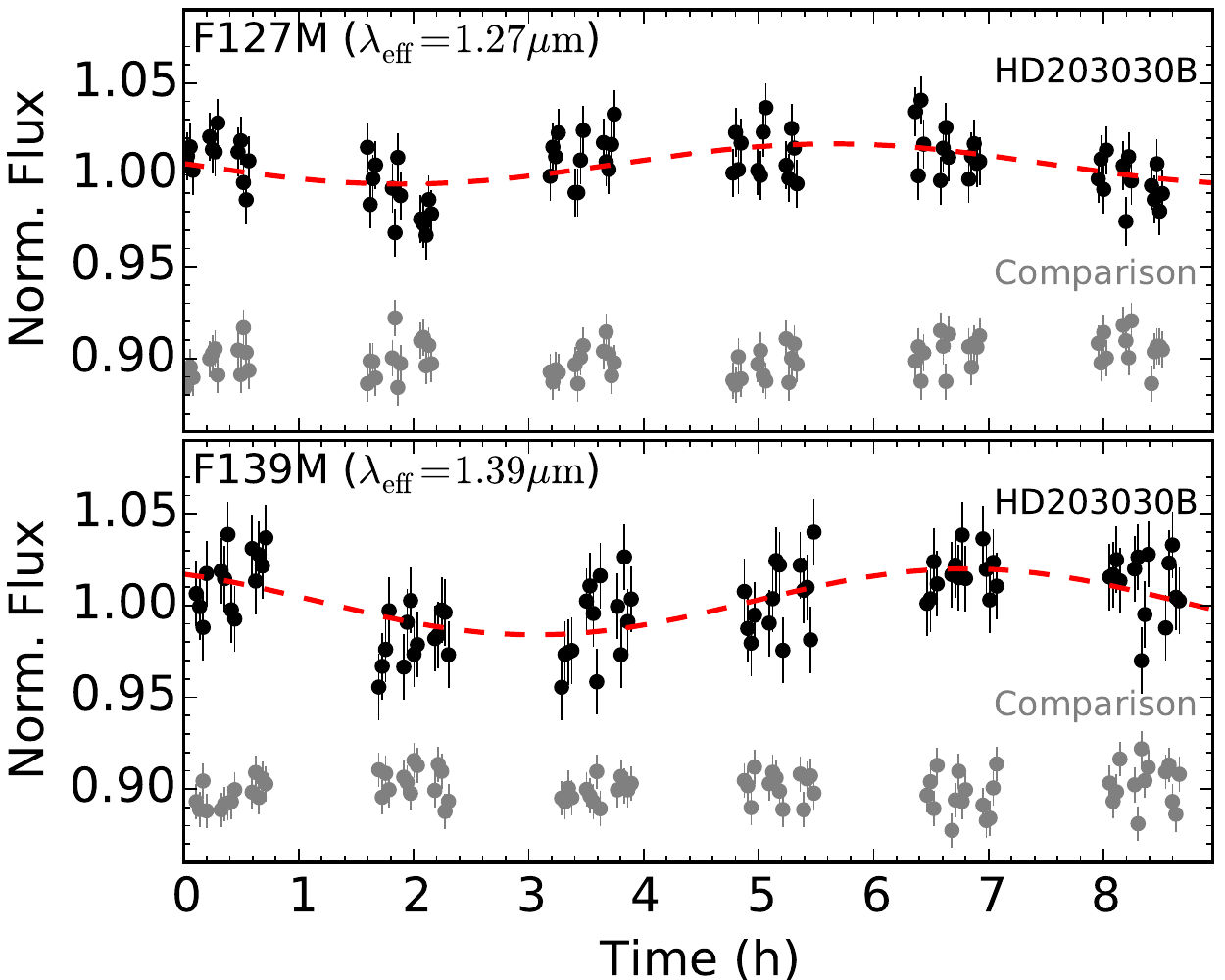}
\caption{Normalized light curves of HD 203030B (black symbols) at 1.27$\mu$m (top) and 1.39$\mu$m (bottom). The normalized light curves of comparison star R2 (see Figure \ref{im}) are also shown (grey symbols) and offset vertically for clarity. R2 is comparable in brightness to HD~203030B and is the brightest comparison star in our field of view. HD~203030B exhibits variability at both wavelengths that is not seen in the comparison star. The dashed curves show the best fit sine curve from the MCMC analysis described in Section \ref{results}. The Modified Julian Date at time zero is 58056.276856.  \label{sine}}
\end{figure}

We measured the fluxes of HD 203030B and 6 other comparison stars (Figure \ref{im}, bottom right) by using aperture photometry and the task {\sc phot} within the Image Reduction and Analysis Facility software (IRAF). For each object, we tried a set of circular apertures with radii in the range 1--6 pixels and sky annuli with an inner radius of 6 pixels and a width of 3 pixels. To retrieve the light curve of HD~203030B, we divided its flux by the total flux of the reference stars, and normalized by the average value of the entire light curve. Photometric uncertainties were obtained from a weighted combination of the individual uncertainties of each star, as derived from the WFC3 detector parameters and photon statistics. The median uncertainties for HD~203030B at F127M and F139M were 1.0\% and 1.5\%, respectively. We compared different combinations of apertures and reference stars, and found that the smallest photometric scatter in both filters was obtained with an aperture of 2 pixels and using reference stars R1, R2, and R3.  The F127M and F139M light curves of HD~203030B and of reference star R2 are shown in Figure \ref{sine}.

\subsection{Near-infrared Spectroscopy \label{specsec}}

We obtained $R\approx130$ 1.07--1.70 $\mu$m spectra of HD~203030B with the G141 grism over two consecutive {\sl HST} orbits on 12 September 2015. These data were too contaminated by scattered light from the HD~203030 primary to be useful for spectrophotometric variability analysis.  We nevertheless present them here in case an improved high-contrast spectropscopic reduction is possible in the future.

\begin{figure}
\centering
\includegraphics[width=0.48\textwidth]{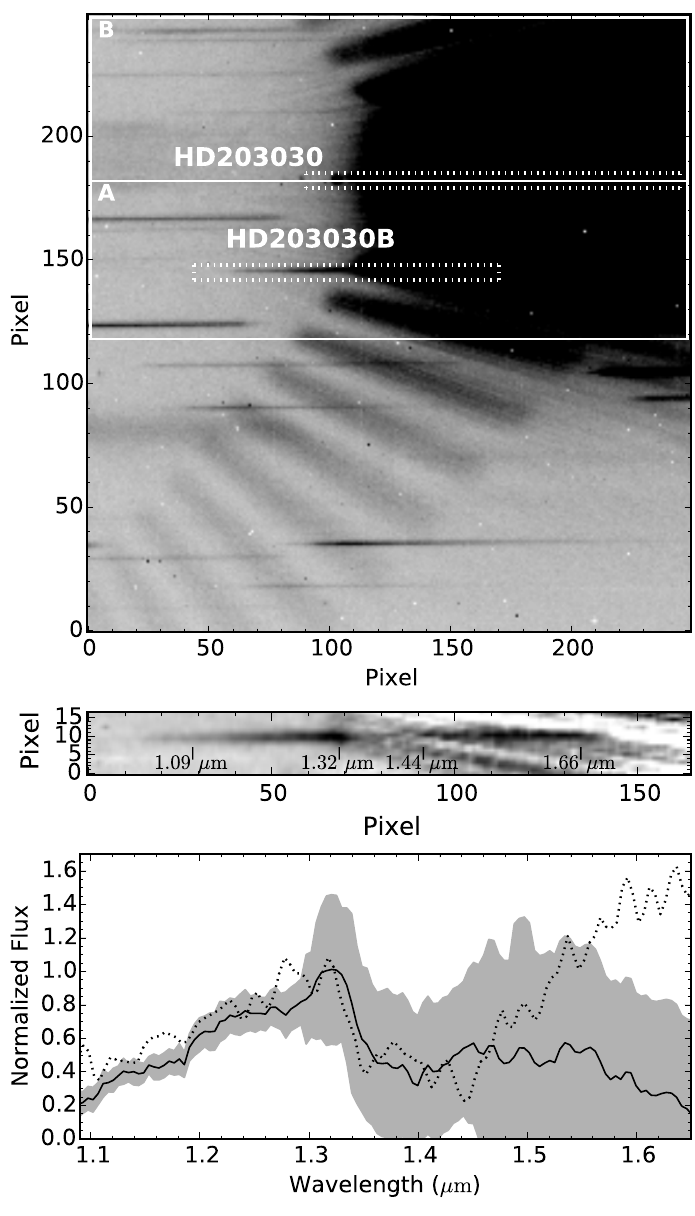}
\caption{{\sl Top:} Raw 2D WFC3 spectrum of HD~203030A and B.  The locations of the spectra of the primary and the secondary are enclosed in white dotted rectangles.  The larger solid rectangle regions `A' and `B' are positioned symmetrically around the trace of the primary, and were used for background subtraction.  Region B was mirrored around the trace of the primary and subtracted from region A, using different scaling factors and offsets (see Section \ref{specsec}). {\sl Middle:} Zoom onto the location of the HD~203030B spectrum after the subtraction of the bright halo of the primary. {\sl Bottom:} Average spectrum of HD 203030B (black) in the 1.10--1.65\,$\mu$m range after combining the 22 individual background-subtracted spectra. The grey area is the standard deviation spanned by the individual spectra. A ground-based IRTF/SpeX spectrum of HD 203030B with similar resolution \citep[dotted;][]{2017AJ....154..262M} is shown for comparison. Both spectra are normalized to their average value in the  1.32--1.33 $\mu$m range.\label{spec}}
\end{figure}

At each orbit, we collected 2--4 images using the filter F132N to establish an accurate position of HD~203030B on the detector, and then obtained 11 spectra with individual exposure times of 201.4 s. A raw image of a single slitless spectrum is shown in the top panel of Figure \ref{spec}. The signal from HD~203030B is heavily contaminated by the saturated halo of the primary, despite the 12$\arcsec$ angular separation. We attempted to correct this contamination using the provided .flt files. First, we corrected bad pixels as done in Section \ref{sec}. Second, we fit for the center of the HD~203030A trace, which we then  used to define two regions positioned symmetrically around the trace, labeled as A and B in the top panel of Figure \ref{spec}. We mirrored region B around the trace and subtracted it from region A. We weighted the mirrored region B by different factors in the range 0.7--1.3 and applied vertical and horizontal offsets to try to improve the subtraction. From this family of differenced images, we chose the one with the smallest standard deviation in the residuals above and below the trace of HD~203030B.  We note that in all trial cases the standard deviation of the residuals in the contaminated area ($\gtrsim$1.32\,$\micron$) was about 10 times larger than that in a non-contaminated region. A background-subtracted image of the HD~203030B spectrum centered on the object's trace is shown in the middle panel of Figure \ref{spec}.

To perform the spectral extraction and wavelength calibration from each of the 22 background-subtracted images, we first expanded the images from the subarray mode to full-frame mode by using a custom Python routine \citep{2013ApJ...768..121A}, and then we used the {\sc aXe} pipeline \citep{2009PASP..121...59K} that works only with full-frame images. The trace of the HD~203030B spectrum in each image is faint, so we extracted it by using a 4 pix-wide aperture without any sky subtraction, as we have already subtracted the local background.  Larger extraction apertures introduced too much noise from surrounding regions. The typical signal-to-noise ratio (SNR) of an individual spectrum of HD~203030B was $\sim$9 at 1.27$\mu$m, which prevented us from investigating the presence of flux changes over the two {\sl HST} orbits.  Thus, we only combined the 22 individual spectra, and show this final spectrum and its associated standard deviation in the bottom panel of Figure \ref{spec} (black). We also show a flux-calibrated spectrum of HD~203030B (dotted line) with a similar resolution as that taken with WFC3, but obtained with IRTF/SpeX from the ground \citep{2017AJ....154..262M}. Both spectra share similar shapes in the 1.1--1.3 $\mu$m range, but differ at the longer wavelengths: because of residual contamination from the halo of the bright primary. WFC3 slitless grism spectra of more widely separated faint companions to bright stars have been successfully extracted in the past \citep[e.g.][]{2018AJ....155..132Z}. However, the HD~203030B data presented here are an extreme case. Hence, we do not consider the {\sl HST} spectrum in the remainder of our analysis.

\begin{figure*}
\centering
\includegraphics[width=0.98\textwidth]{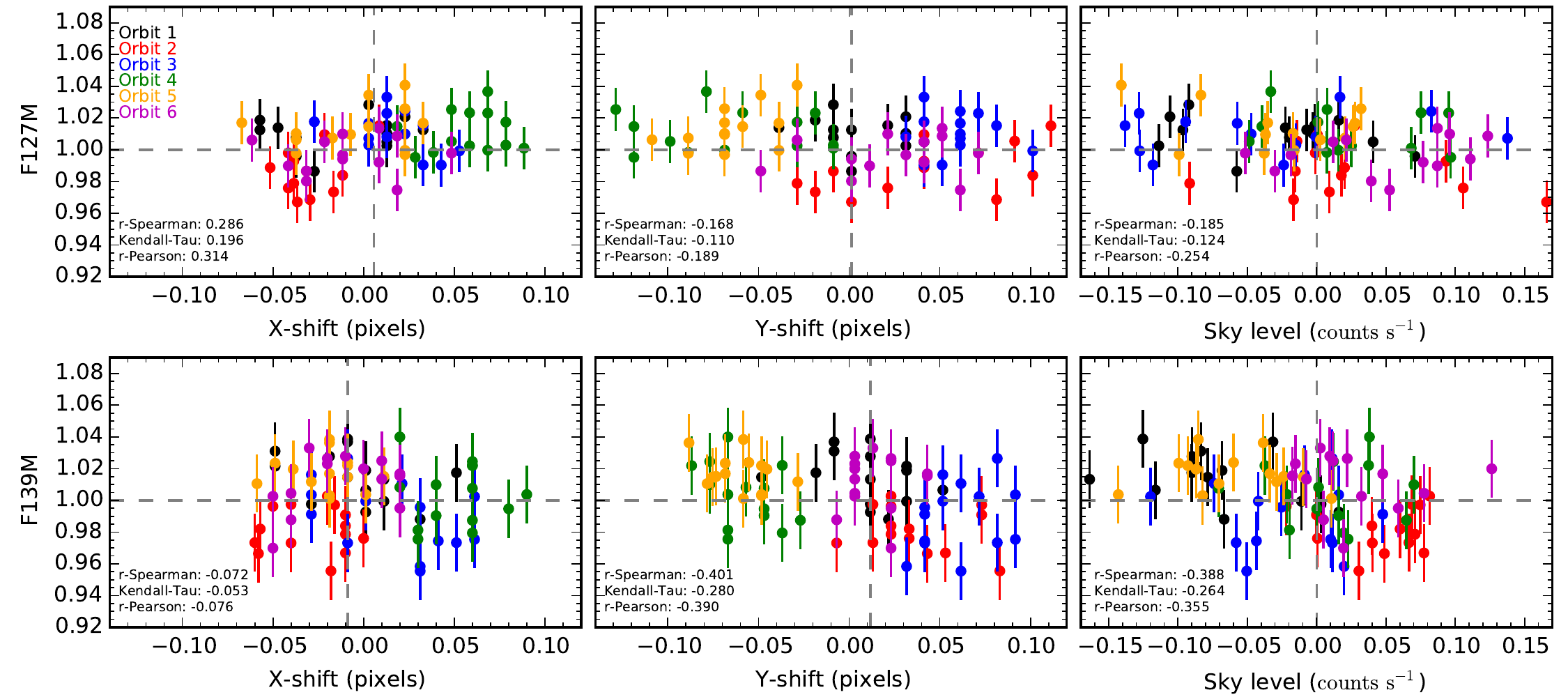}
\caption{Normalized fluxes of HD 203030B (F127M top, F139M bottom) as a function of the $x$ or $y$ centroid positions on the detector (left, middle), or of the sky level (right). No clear correlation is evident between these parameters and the flux of HD 203030B. Grey, dashed lines indicate mean values. The values for the Pearson's $r$, Spearman's $r$, and Kendall's $\tau$ correlation coefficients are also indicated. The photometry from each orbit is depicted with a different color. \label{sys}}
\end{figure*}

\section{Results and Analysis}\label{results}
\subsection{HD 203030B is Periodically Variable over 1.2--1.4 $\mu$m}
 
 We detect flux changes in the photometric light curves of HD 203030B that are not seen in the light curve of comparison star R2  (Figure \ref{sine}), or in the sum of the comparison star fluxes. We investigated whether this modulation could be induced by  measurement systematics, such as the position of the stellar centroid in each image, or the background brightness at the object position. The normalized flux of HD~203030B during each orbit vs.\ each of these parameters is shown in Figure \ref{sys}. No clear correlation is seen. To test this we computed Pearson's $r$, Spearman's $r$, and Kendall's $\tau$ coefficients of correlation, that we also show in Figure \ref{sys}. Their values are small, indicating no significant statistical correlation of the normalized flux of HD 203030B with the object's position on the detector or with the sky level. This is expected because if there were some systematics affecting the differential photometry, they should also induce a similar shape in the light curve of the comparison star, which is however flat (Figure~\ref{sine}).

\begin{figure}
\centering
\includegraphics[width=0.49\textwidth]{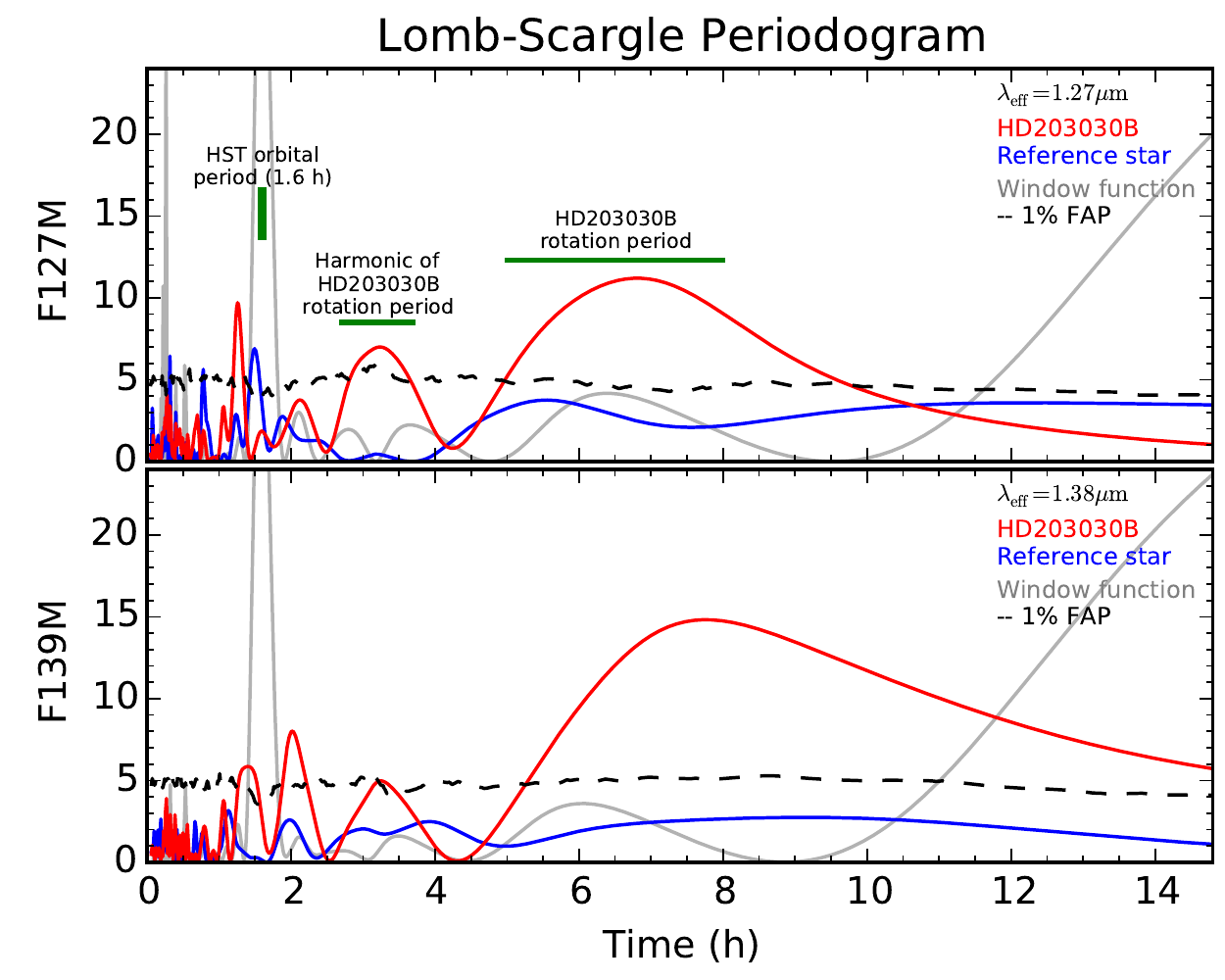}
\caption{LS periodogram of the light curves of HD 203030B (red), reference star R2 (blue), and the window function (grey) at 1.27$\mu$m (top) and 1.39$\mu$m (bottom). The dashed line indicates the 1\% false-alarm-probability, calculated from 10$^4$ simulated light curves using our data and their associated uncertainties. Relevant peaks in the periodogram of HD~203030B (see Section \ref{results}) are also indicated.
\label{per}}
\end{figure}

The main result from Figures \ref{sine} and \ref{sys} is that HD 203030B exhibits temporal variability at both 1.27 $\mu$m and 1.39 $\mu$m that: $i)$ is not seen in other comparison stars in the same field of view, and $ii)$ is not attributable to instrumental systematics. Hence, we searched for periodicity in both light curves. Figure \ref{per} shows the Lomb-Scargle periodogram \citep{1976Ap&SS..39..447L,1982ApJ...263..835S} corresponding to the light curves of HD 203030B (red), comparison star R2 (blue), and the window function (grey), related to the sampling of our data. The periodograms of HD 203030B in both filters display a significant peak in the 6--8 h range that is not seen in the periodogram of the comparison star. We also find a significant peak in both filters close to $\sim$3.5 h that we interpret as a harmonic of the primary peak seen at 6--8 h. Finally at 1.5--2 h there are other significant peaks present in the periodograms of HD~203030B and the comparison star that we attribute to the {\sl HST} orbital period (1.6 h), as they are also seen in the periodogram of the window function. 

\begin{figure*}
\centering
\includegraphics[width=0.99\textwidth]{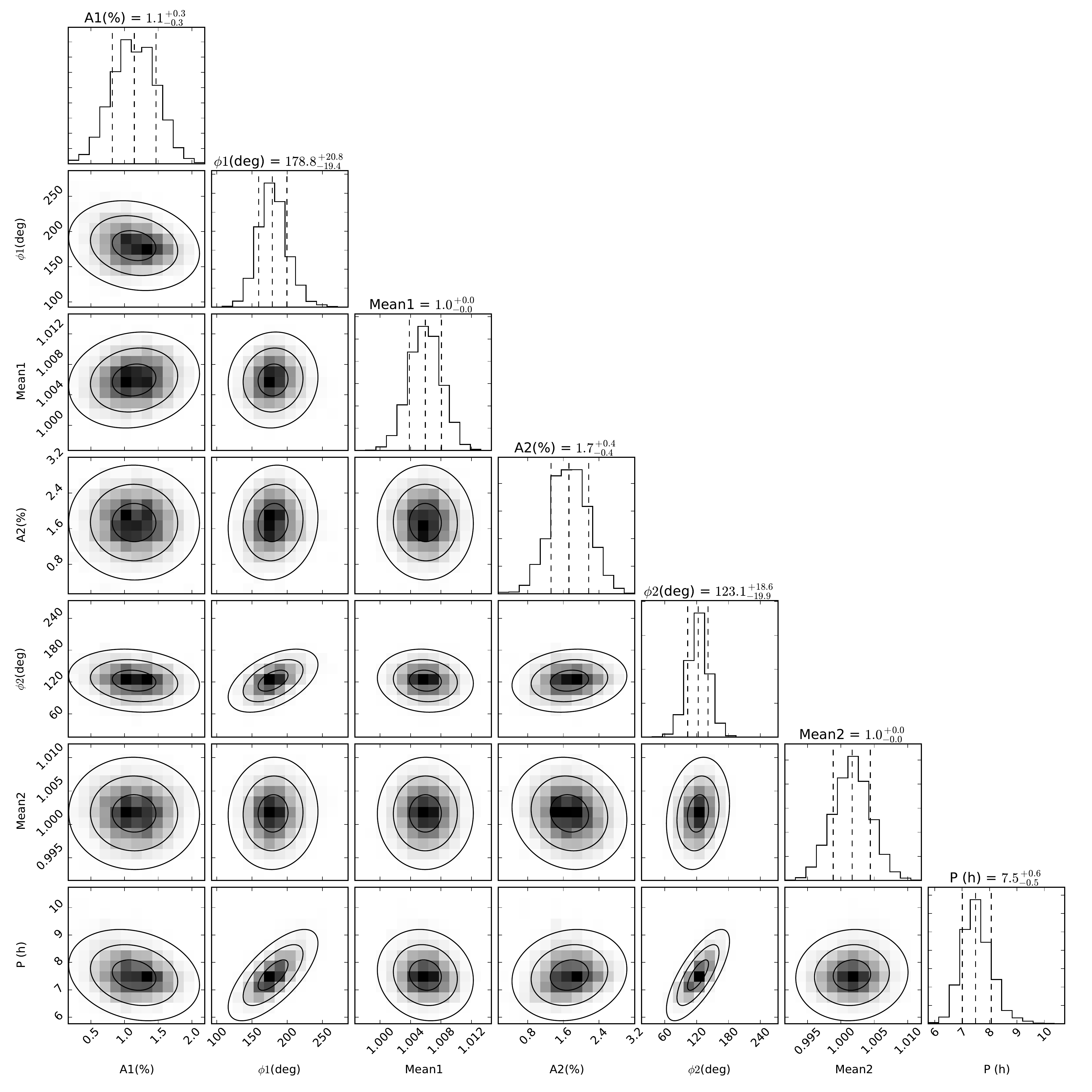}
\caption{Posterior distributions of the light curve parameters from our MCMC sine curve fits to the HD~203030B photometry. Indices 1 and 2 denote data at 1.27\,$\mu$m and 1.39\,$\mu$m, respectively. We allowed the amplitudes ($A$), phases ($\phi$), and zero levels ($K$) to vary uniformly over 0\%--10\%, 0--2$\pi$, and 0.9--1.1, respectively, while the period ($P$) was fixed to be the same for both filters. Vertical dashed lines indicate the 16 and 84 percentiles and the mean value of the distributions. The ellipses show the 1$\sigma$, 2$\sigma$, and 3$\sigma$ credible intervals. Best fit values are shown in Table \ref{tab1}. \label{mc}}
\end{figure*}

Independently from the output of the periodogram, we used a Markov Chain Monte Carlo routine \citep[MCMC;][]{2005blda.book.....G} to fit sine functions simultaneously to both HD~203030B light curves. We opted for a sine, i.e., $F\,=A\,{\rm sin(}2\pi\,t/P + \phi{\rm )}+K$, since it is the simplest periodic function, and because the light curves do not exhibit a more complicated shape.  A sine curve is an adequate representation of a rotational modulation caused by a planetary-scale wave, as may be common in brown dwarfs \citep{2017Sci...357..683A}. We assumed that the photometric period ($P$) is the same in both filters and takes values in the range 1--40 h (typical range seen in most ultra-cool dwarfs), while the amplitude ($A$), phase ($\phi$) and the mean level ($K$) can vary independently in each filter in the ranges 0\%--10\%, 0--2$\pi$, and 0.9--1.1, respectively. We used flat priors to sample the parameter space and performed $5\times10^5$ iterations in our chain (the first $5\times10^4$ were discarded as part of the burn-in stage). Our results for the posterior distribution of the variables are shown in Figure \ref{mc} and summarized in Table \ref{tab1}. We find amplitudes of variability of $A_{1.27\mu{\rm m}}=1.1\%\pm0.3\%$ and $A_{1.39\mu{\rm m}}=1.7\%\pm0.4\%$, and a period of $7.5^{+0.6}_{-0.5}$ h (1$\sigma$). The best fit derived from this analysis is shown in Figure \ref{sine} (dashed line). The period obtained from our MCMC analysis is in agreement with the primary peak at 6--8 h identified in the periodogram analysis. At 1$\sigma$, this period is compatible with being a multiple of the {\sl HST} orbital period (i.e., 1.6 h $\times$\,5). However, this is unlikely as we do not recover a similar period value when repeating the MCMC analysis for the comparison star. The posterior distribution for the period obtained from the data of the comparison star is flat, and does not show any preferred value in the investigated phase space. Our data only cover $\sim$1.1 rotations of the period found from our MCMC analysis, and so further observations could refine the measured periodicity. 

The retrieved HD~203030B period is in the same range as those seen in other young objects \citep{2015ApJ...809L..29S,2016ApJ...818..176Z,2016ApJ...829L..32L,2017Sci...357..683A,2018AJ....155...95B,2018MNRAS.474.1041V,2018AJ....155...11M,2018AJ....155..132Z}.  It conforms with the finding of \citet[][]{2018AJ....155..238S} that the median rotation period of 10--300 Myr brown dwarfs is about 10 hr, more than twice the value of the median rotation period of field-age brown dwarfs ($\sim$4 hr).

\begin{table}[h!]
\renewcommand{\thetable}{\arabic{table}}
\centering
\caption{Rotational modulation of HD 203030B from our MCMC analysis.}
\label{tab1}
\begin{tabular}{lcc}
\tablewidth{0pt}
\hline
\hline
 \multicolumn{3}{c}{$F\,=A\,{\rm sin(}2\pi\,t/P + \phi{\rm )}+K$}\\
 \hline
 & F127M & F139M   \\
\hline
 $A$ ($\%$)& $1.1\pm0.3$ & $1.7\pm0.4$\\
 $\phi$ (deg)&$179\pm20$ & $123\pm19$ \\
 $K$ & $1.000\pm0.002$ & $1.000\pm0.003$\\
 $P$ (h)& \multicolumn{2}{c}{$7.5^{+0.6}_{-0.5}$} \\
\decimals
\hline 
\end{tabular}
\tablenotetext{}{Modified Julian Date for zero time is 58056.276856.}
\raggedright
\scriptsize
\end{table}

\subsection{The 1.27--1.39$\mu$m  rotational modulation of HD~203030B is Likely Wavelength-dependent}
\label{sec:wavelength_dependence}

We chose the F127M and F139M filters for the WFC3 observations to sample differences between the water-free 1.27\micron\ region and the water-absorbed 1.39\micron\ region. We find marginal ($\sim$2$\sigma$) differences in both the amplitudes and the phases of the F127M and F139M light curves.  If real, these would disagree with the linear amplitude and constant phase behavior across 1.1--1.7~$\mu$m wavelengths observed by \citet{2018AJ....155...11M} in the somewhat older L6  dwarf LP~261--75B \citep[$\approx$300~Myr;][]{liu_etal16}, or in field mid-L dwarfs \citep{2016ApJ...826....8Y}.  

However, similar wavelength-dependent 1.1--1.7~$\mu$m rotational modulation is seen in other young, cool late-L dwarfs.  The planetary-mass L7 member of the $\approx$23~Myr \citep{mamajek_bell14} $\beta$ Pictoris moving group PSO~J318.5--22 \citep{2013ApJ...777L..20L} has a smaller (2.38\%) semi-amplitude in the 1.34--1.44~$\mu$m water absorption region compared to the broadband $J$ filter \citep[2.92\%;][]{2018AJ....155...95B}.  The L6.5 member of the $\approx$125~Myr\citep{barenfeld_etal13} AB~Doradus moving group WISEP J004701.06+680352.1 \citep{2012AJ....144...94G, 2015ApJ...813..104G} also shows diminished variability between 1.34--1.44~$\mu$m, compared to the trend in the surrounding continuum \citep{2016ApJ...829L..32L}.  PSO~J318.5--22 and WISEP J004701.06+680352.1 are the only two other young late-L dwarfs, besides HD~203030B, to have received precise 1.1--1.7$\micron$ spectrophotometric monitoring with HST/WFC3.

One issue that sets the variability of HD~203030B apart from that of PSO~J318.5--22 and WISEP J004701.06+680352.1 is that the amplitude in the F139M water band is marginally \textit{higher} than in the shorter-wavelength F127M continuum.  In addition, neither PSO~J318.5--22 nor WISEP J004701.06+680352.1 show phase shifts in the 1.1--1.7$\mu$m region.  As our evidence for differences in the F127M and F139M amplitudes and phases is at the $\approx$2$\sigma$ level only, it is possible that HD~203030B behaves consistently with other young late-L dwarfs.  Alternatively, the flux variations may be driven by thermal perturbations deeper in the atmosphere.  Time-dependent simulations of the atmospheric thermal structure under such perturbations show that they can produce wavelength-dependent variability amplitudes and phase shifts in high-gravity T dwarfs \citep{robinson_marley14}.  The dependence of the effect of such thermal perturbations on warmer or lower-gravity ultra-cool atmospheres has yet to be explored.

We assess the broader evidence for wavelength-dependent spectrophotometric variations in L dwarfs below.

\subsection{Wavelength-dependent 1.1--1.7$\mu$m Variability in Young Late-L Dwarfs Points to Condensate Sedimentation}

Precise spectrophotometric variability monitoring of L and T dwarfs with HST/WFC3 is enabling a comparison of their water vapor and cloud condensate content over the corresponding range of atmospheric pressures.  The \textit{Cloud Atlas} program has shown that $\gtrsim$300~Myr-old L dwarfs exhibit at most a linear wavelength dependence of variability over 1.1--1.7$\mu$m, while T dwarfs are significantly less variable in the 1.34--1.44~$\mu$m water absorption band compared to the surrounding continuum \citep{2016ApJ...829L..32L,2018AJ....155...11M, 2018AJ....155..132Z, apai_etal19}. This dichotomy is explained by the positioning of the highest condensate cloud layer relative to the top of the water vapor column \citep{2015ApJ...798L..13Y}. The cloud layer resides above the water vapor column in field L dwarfs, and so solely determines their 1.1--1.7$\mu$m variability amplitudes and phases.  Field L dwarf variability across this wavelength range is thus fully phased, and nearly wavelength-independent.  In cooler T dwarfs the condensate clouds have sunk below the top of the water vapor column, and so the influence of the clouds on the variability in the high-altitude 1.34--1.44$\mu$m water vapor band is diminished \citep{2013ApJ...768..121A, apai_etal19}.  Consequently, the variability characteristics of T dwarfs in the 1.34--1.44$\mu$m water band, which probes low-pressure high-altitude atmospheric layers, become decoupled from those in the surrounding continuum, which probes deeper in the atmosphere.

Unlike their older counterparts, $\lesssim$150~Myr-old late-L dwarfs appear to show a similar altitude differentiation between the top condensate cloud layer and the top of the water vapor column as do early T dwarfs (Sec.~\ref{sec:wavelength_dependence}).  The explanation is likely rooted in a known similarity between young late-L dwarfs and T dwarfs.  Young L/T-transition dwarfs are 100--300~K cooler than their $\approx$1300~K field-age counterparts \citep{2006ApJ...651.1166M, luhman_etal07b, dupuy_etal09a, 2016ApJS..225...10F}. Young late-Ls nevertheless lack methane absorption in the near-infrared because of a significant departure from CO/CH$_4$ chemical equilibrium, and an accordingly diminished methane abundance at low surface gravity \citep{barman_etal11, barman_etal11b}. The cooler temperatures of young late-Ls would enhance condensate growth, while their lower surface gravities could increase the altitude differentiation between the (denser) condensates and the top of the (less dense) water vapor column.  The result would be a low-gravity L-type spectrum with a T dwarf-like decoupling of the variability characteristics in and out of the water band, as observed.

The wavelength-dependent nature of the 1.1--1.7 $\mu$m variability of young late-L dwarfs is a new manifestation of the role of surface gravity at the L/T transition, adding to its known effect on effective temperature and luminosity.

\citet{marley_etal12} offer a theoretical picture on the effect of decreased surface gravity on cloud formation and sedimentation at the L/T transition.  Their construct accounts for the survival of photospheric clouds to lower effective temperatures at lower surface gravities, as is inferred from the characteristically red spectral energy distributions of young late-L dwarfs.  \citet{marley_etal12} conclude that for two brown dwarfs of the same effective temperature, the condensate clouds reside higher in the atmosphere of the lower surface gravity brown dwarf.  

We compare this theoretical picture to our synthesis of late-L dwarf spectrophotometric variability from {\sc HST/WFC3}.  We find that for two late-L dwarfs of the same spectral subtype (e.g., L7), the younger one with lower gravity exhibits significant wavelength-dependent differences in and out of the water band, whereas the older one does not.  In the context of cloud heights, we conclude that the positioning of the condensate clouds relative to the level where the water column becomes opaque is \textit{lower} in the low-gravity late-L dwarf.  However, these two constructs do not disagree: precisely because young late-L dwarfs have lower effective temperatures at the same spectral subtype.  Conversely, at the same $\lesssim$1100~K effective temperatures, young late-L dwarfs show comparable or smaller variability amplitudes in 1.1--1.7$\micron$ HST/WFC3 monitoring \citep[][this paper]{2016ApJ...818..176Z, 2016ApJ...829L..32L, 2018AJ....155...95B} than field early-to mid-T dwarfs \citep{2013ApJ...768..121A, 2015ApJ...798..127B, 2016ApJ...826....8Y, 2018AJ....155..132Z}: in full agreement with the \citet{marley_etal12} picture.

\subsection{Implications for Low-gravity L Dwarfs and Young Giant Planets}

In the preceding construct, flux variability in wavelength regions outside of major sources of gas opacity is driven by condensate hazes or clouds of non-uniform thickness that reside at deeper, hotter atmospheric layers.  While this is the accepted explanation for T dwarfs, we have posited that a similar differentiation between the scale heights of condensates and water vapor can also exist in low-gravity late-L dwarfs, or more generally, between condensates and any gas species in low-gravity L dwarfs.  

We anticipate that this construct may also hold for earlier-type L dwarfs.  That is, low-gravity early- to mid-L dwarfs may also be experiencing a differentiation between the condensate cloud layer and any gas species, such as water or carbon monoxide vapor.  The consequence would be potentially enhanced variability amplitudes in young L dwarfs in wavelength regions free of high-altitude gas opacity.  Most favorable for detecting such variability would be the 1.20--1.34$\mu$m window, as we have pursued here with the F127M WFC3 filter, since it probes deepest into L-dwarf atmospheres \citep{2001ApJ...556..872A,2016ApJ...826....8Y}.  However, other wavelength regions in between the water and carbon monoxide absorption bands, could also be suitable: such as the {\sl Spitzer} 3.6 $\mu$ band or the near-infrared $J$ band.

There is corroborating evidence that low-gravity L dwarfs indeed exhibit higher-amplitude variations than their older field counterparts at these wavelengths.  \citet{2015ApJ...799..154M} infer this tentatively from \textit{Spitzer} 3.6$\mu$m-band ($\lambda_{\rm central}=3.55\mu$m, FWHM $=0.75\mu$m) monitoring of L3--L5.5 dwarfs.  The \textit{Spitzer} 3.6$\mu$m band is relatively free of major sources of gas opacity in L dwarfs \citep{burrows_etal01}.  Probing deeper into L-dwarf atmospheres with $J$-band monitoring, \citet{2018MNRAS.474.1041V} confirm with 98\% confidence that young L0--L8.5 dwarfs exhibit higher-amplitude variations than field L dwarfs.  

We note as a corollary that if altitude differentiation between condensates and gasses is the reason for the enhanced variability in low-gravity L dwarfs, then the effect may not persist into the T dwarfs, since the condensates in their atmospheres have already sunk below the scale heights of the dominant gas species.

The implication for self-luminous young giant planets, such as the handful that have already been directly imaged, is straightforward.  Young L-type giant planets, if adequately inclined to reveal flux modulations with rotation, would be most variable in the 1.20--1.34$\micron$ region.  At an estimated mass of $\sim$11 times Jupiter's \citep{2017AJ....154..262M}, HD~203030B is one of the few known examples of variable planetary-mass companions, along with 2MASS 1207--3932b \citep{2004A&A...425L..29C, 2016ApJ...818..176Z}, HN~PegB \citep{luhman_etal07b, 2018AJ....155..132Z}, Ross~458C \citep{2019ApJ...875L..15M}, AB~PicB \citep{chauvin_etal05b, 2019AJ....157..128Z}, and 2MASS~0122--2439B \citep{2013ApJ...774...55B, 2019AJ....157..128Z}. More precise high-contrast near-infrared spectrophotometry \citep[cf.,][]{apai_etal16} could reveal variability in closer-in extrasolar giant planets, too.

\section{CONCLUSIONS}\label{conclusions}

We collected imaging and spectroscopic data of the 30--150~Myr-old L7.5 dwarf HD~203030B with the near-infrared channel of WFC3 on the {\sl HST} over 6 and 2 orbits, respectively. The photometric data were collected during each orbit by alternating two filters centered at wavelengths inside (F139M) and outside of (F127M) the 1.34--1.44$\micron$ water band. Both sets of imaging data show clear modulation that cannot be explained by instrumental systematics and that is not seen in the light curves of other stars in the WFC3 field. We found a likely rotation period of $7.5^{+0.6}_{-0.5}$ h and a phase lag of 56$^\circ \pm$28$^\circ$ between the light curves of the two filters. Unfortunately, the spectroscopic data were too contaminated by the halo of the bright primary to assess any rotation-induced variability.

HD~203030B shows marginal evidence for differences in both the variability amplitude and phase between the light curves at water-absorbed (F139M) and water-free (F127M) wavelengths: a behavior not seen in warmer L dwarfs, but common in T dwarfs.  This could be an indication that the patchy cloud layer in this young very late L dwarf resides near or below the level where the water column becomes opaque.  The $T_{\rm eff}=1040\pm50$K effective temperature of HD~203030B is already known to be well below that of older late-L dwarfs \citep{2006ApJ...651.1166M, 2017AJ....154..262M}.  Its low effective temperature could thus facilitate the sinking of the cloud layer responsible for the F127M variations.  Similar wavelength-dependent spectrophotometric variability is also seen in the only two other young and similarly cool late-L dwarfs monitored with HST/WFC3.  
We suggest that this condensate/gas differentiation mechanism could explain the enhanced variability amplitudes of low-gravity L dwarfs or L-type self-luminous giant planets in general, and that their variability will be most pronounced in the water-free 1.20--1.34$\micron$ spectral window.  Because the levels of condensate and gas opacities are already well differentiated in T dwarfs, we do not expect a similar enhancement of the variability amplitudes among young T dwarfs.

\acknowledgments

Based on observations made with the NASA/ESA Hubble Space Telescope, obtained at the Space Telescope Institute, which is operated by AURA, Inc., under NASA contract NAS 5-26555, under GO 14241.  We are grateful for research support provided by the Natural Sciences and Engineering Research Council of Canada (grant no.\ RGPIN-04396-2014) and the Canada Research Chairs Program.

\facility{HST: (WFC3), Gaia.}
\software{IRAF \citep{1986SPIE..627..733T,1993ASPC...52..173T},  Corner plot \citet{2016JOSS....1...24F}.}

\bibliography{biblio,bibliography}
\clearpage

\end{document}